\documentclass[12pt]{article}

\usepackage{times}
\usepackage{graphicx}
\usepackage{amssymb}

\begin{document}

% Double-space the manuscript.
\baselineskip24pt

\begin{center}
% TITLE
{\Huge\bf Electric field control of spin transport}
\end{center}

\begin{center}
% AUTHORS
  {\large Sangeeta Sahoo$^{\ast}$, Takis Kontos$^{\ast}$, J{\"u}rg Furer, Christian Hoffmann \\
  Matthias Gr\"{a}ber, Audrey Cottet \& Christian Sch\"{o}nenberger}

  Institut f\"{u}r Physik, Universit\"{a}t Basel, Klingelbergstrasse 82, CH-4056 Basel, Switzerland. \\
  $^\ast$These authors contributed equally to this work \\
  \normalsize{Correspondence should be addressed to C.S.; E-mail: Christian.Sch{\"o}nenberger@unibas.ch}
\end{center}

\date{}

% ===============================================================================================
% END of PREAMPLE
% ===============================================================================================

% ===============================================================================================
% ABSTRACT
% ===============================================================================================
{\bf Spintronics is an approach to electronics in which the spin
of the electrons is exploited to control the electric resistance
$R$ of devices~$^{1-3}$. One basic building block is the
spin-valve~$^{4-6}$, which is formed if two ferromagnetic
electrodes are separated by a thin tunneling barrier. In such
devices, $R$ depends on the orientation of the magnetisation of
the electrodes. It is usually larger in the antiparallel than in
the parallel configuration. The relative difference of $R$, the
so-called magneto-resistance (MR), is then \emph{positive}.
Common devices, such as the giant magneto-resistance sensor
used in reading heads of hard disks~$^{7,8}$, are based on this
phenomenon. The MR may become anomalous (\emph{negative})~$^{9}$,
if the transmission probability of electrons through the device is
spin or energy dependent~$^{3}$. This offers a route to the
realisation of gate-tunable MR devices~$^{10,11}$, because
transmission probabilities can readily be tuned in many devices
with an electrical gate signal. Such devices have, however, been
elusive so far. We report here on a pronounced gate-field
controlled MR in devices made from carbon nano\-tubes with
ferromagnetic contacts. Both the amplitude \emph{and} the sign of
the MR are tunable with the gate voltage in a predictable manner.
% We emphasise that this spin-field effect is not restricted to carbon
% nano\-tubes but constitutes a generic effect which can in
% principle be exploited in all resonant tunneling devices.
} % end abstract

\newpage
% ===============================================================================================
% END of ABSTRACT
% ===============================================================================================

% ===============================================================================================
% START of DOCUMENT
% ===============================================================================================

% In setting up this template for *Science* papers, we've used both
% the \section* command and the \paragraph* command for topical
% divisions.  Which you use will of course depend on the type of paper
% you're writing.  Review Articles tend to have displayed headings, for
% which \section* is more appropriate; Research Articles, when they have
% formal topical divisions at all, tend to signal them with bold text
% that runs into the paragraph, for which \paragraph* is the right
% choice.  Either way, use the asterisk (*) modifier, as shown, to
% suppress numbering.

% \section*{Introduction}

Early work on spin transport in multi-wall carbon nano\-tubes
(MWNTs) with Co contacts showed that spins could propagate
coherently over distances as long as \mbox{$250$\,nm}~$^{12}$. The
tunnel magneto-resistance $TMR=(R_{AP}-R_{P})/R_{P}$, defined as
the relative difference between the resistances $R_{AP}$ and
$R_{P}$ in the antiparallel (AP) and parallel (P) magnetisation
configuration, was found to be positive and amounted to
\mbox{$+4$\,\%} in agreement with Julli\`ere's formula for tunnel
junctions~$^{4,13}$. A negative $TMR$ of about \mbox{$-30$\,\%}
was reported later for MWNTs contacted with similar Co
contacts~$^{14}$. In these experiments, the nano\-tubes did not
exhibit a quantum dot (Qdot) behaviour. It has been shown,
however, that single-wall carbon nano\-tubes (SWNTs) and MWNTs
contacted with non-ferromagnetic metals could behave as Qdots and
Fabry-Perot resonators~$^{15-19}$, in which one can tune the
position of discrete energy levels with a gate electrode. From
this, one can expect to be able to tune the sign and the amplitude
of the $TMR$ in nano\-tubes, in a similar fashion as predicted
originally for semiconductor heterostructures~$^{11}$.

In this letter, we report on $TMR$ measurements of multi-wall and
single-wall carbon nano\-tubes that are contacted with ferromagnetic electrodes
and capacitively coupled to a back-gate~$^{20}$.
A typical sample geometry is shown in the inset of Fig.~1.
As a result of resonant tunneling, we observe a striking oscillatory
amplitude and sign modulation of the $TMR$ as a function of the gate voltage.
We have studied and observed the $TMR$ on $9$ samples ($7$ MWNTs and $2$ SWNTs)
with various tube lengths $L$ between the ferromagnetic electrodes (see methods).
We present here results for one MWNT device and one SWNT device.
% for which the spacings between the ferromagnetic
% electrodes were $L=0.4$ and \mbox{$0.5$\,$\mu$m}, respectively.

We first discuss the results of the MWNT device. Figure~1 displays
single traces of the linear response resistance $R$ as a function
of the magnetic field $H$ at \mbox{$1.85$\,K} for two sweep
directions and four different gate voltages $V_g$. For all cases,
the characteristic hysteretic behaviour of a spin valve appears.
Upon sweeping the magnetic field from \mbox{$-500$\,mT} to
\mbox{$500$\,mT}, the configuration gets antiparallel (AP) between
\mbox{$0$\,mT} and \mbox{$100$\,mT}, whereas it is always parallel
(P) for \mbox{$|H|> 100$\,mT}. At \mbox{$V_{g}=-3.1$\,V} for example,
$R$ increases from $49.7$ to
\mbox{$51.5$\,k$\Omega$} when the sample switches from the P to
the AP configuration. This yields a normal \textit{positive} $TMR$
of \mbox{$+2.9$\,\%}.
% \cite{TMRvalue}.
In contrast, at \mbox{$V_{g}=-3.3$\,V}, $R$
switches from \mbox{$30.5$\,k$\Omega$} in the P configuration to a
\textit{smaller} resistance of \mbox{$29.5$\,k$\Omega$} in the AP
configuration, yielding an anomalous \textit{negative} $TMR$ of
\mbox{$-3.5$\,\%}. Therefore, the \textit{sign} of the $TMR$
changes with the gate voltage, demonstrating a
gate-field tunable magnetoresistance (MR).

Figure~2a displays the variation of the $TMR$ in a large $V_g$
window of \mbox{$-5\dots 2$\,V} at $T=1.85$\,K. The $TMR$
is observed to oscillate relatively regularly between
\mbox{$-5$\,\%} and \mbox{$+6$\,\%} on a gate-voltage scale of
\mbox{$\Delta V_g^{TMR}= 0.4 - 0.75$\,V}.
% Taking the visible substructure of the $TMR$ modulation in regions without sign
% change into account, yields a slightly smaller characteristic gate
% voltage scale of \mbox{$\Delta V_{g}^{TMR}\approx 0.4$\,V}.
Two possible mechanisms may account for oscillations in spin
transport: quantum interference~$^{11}$ and gate-field induced
spin-precession via the Rashba spin-orbit interaction~$^{21}$
proposed by Datta and Das~$^{10}$. In the latter case, the spin
orbit interaction yields a spin precession which is reflected both
in the $TMR$ and the conductance.
% Egger and De Martino have shown that subband
% mixing, which can be relevant in MWNTs, does not spoil in
% principle the Datta-Das mechanism thanks to the particular
% structure of the spin-orbit interaction in nano\-tubes \cite{Egger:05}.
To lowest order, the spin-precession would lead to a
$TMR\propto\cos(2mL\beta_R eV_g/\hbar^2)$, where $L$ is the length
of the MWNT, $m$ ($e$) the electron mass (charge), and $\beta_R$
the Rashba spin-orbit parameter. However, the measured magnitude
of \mbox{$\Delta V_{g}^{TMR}\sim 1$\,V} for one $TMR$ period
requires a large \mbox{$\beta_R\sim 10^{-12}$\,m}. Although a
$\beta_R$ value of this order of magnitude has been reported in
semiconductor heterostructures~$^{22}$, this value is unreasonably
large for carbon nano\-tubes, due to the small spin-orbit
interaction of carbon leading to an electron $g$-factor close to
$2$~$^{19,23,24}$. We therefore conclude that the mechanism of
$TMR$ oscillations is quantum interference instead. To
substantiate this, we compare next the $TMR$ gate-voltage scale
$\Delta V_{g}^{TMR}$ with the corresponding scale $\Delta
V_{g}^{e}$ for the addition of single electrons.

To resolve single-electron states, the same sample
was measured at lower temperatures, i.e. at \mbox{$T=300$\,mK}
in a different cryostat which is not suitable for $TMR$ studies.
A measurement of the differential conductance $dI/dV$ as a function of
source-drain $V_{sd}$ and  gate voltage $V_{g}$ at
\mbox{$T=300$\,mK} in a relatively narrow $V_g$ range
is shown in Fig.~2b.
It displays the diamond-like pattern characteristic for
single-electron tunneling in a Qdot.
The visible diamonds vary in size with single electron addition
energies ranging between $0.5$ and \mbox{$0.75$\,meV}, in agreement
with previous reports on MWNT Qdots with
non-ferromagnetic leads~$^{19}$.
% \cite{19,Buitelaar:02b}.
Comparing with $V_{g}^{TMR}$, we see that the average gate-voltage scale
\mbox{$\Delta V_g^{e}=25$\,mV} associated to one single particle level is
much smaller than \mbox{$\Delta V_{g}^{TMR}$}.
The latter corresponds therefore to the addition of at least $16$
electrons rather than $1$.

A gate-voltage scale which agrees with
the $TMR$ signal becomes visible, if the linear conductance $G$ at
low temperatures is monitored over a wider gate-voltage range.
This is shown in Fig.~2c. The single-electron
conductance peaks are strongly modulated in amplitude, leading to
a regular beating pattern with the proper gate-voltage scale
of \mbox{$\Delta V_g \approx 0.4$\,V}.
Note, the absolute values of $V_g$ cannot be compared with
the $TMR$ measurement in Fig.~1a, because the sample was
thermally cycled.

% In order to see, whether a band-structure effect may account for
% the beating we evaluate the respective energy-scales. Each beating
% is made up of a bunch of $\sim 16$ single particle peaks. Taking
% the measured addition energy, which on average is
% \mbox{$0.6$\,meV}, $\Delta V_g^{TMR}$ corresponds to
% \mbox{$\Delta E = 9$\,meV}. This value is an upper bound for the gate-voltage
% induced shift of the chemical potential of the nano\-tube, because
% the Coulomb energy has not been subtracted. The mean spacing
% between one-dimensional subbands $\Delta E_{sb}$ is given by
% $\hbar v_F/d$, where $d$ is the diameter of the nano\-tube and
% $v_F$ the Fermi velocity. Even for a relatively large nano\-tube
% with \mbox{$d = 20$\,nm}, $\Delta E_{sb}$ amounts to
% \mbox{$33$\,meV}, substantially larger than $\Delta E$. This rules
% out a simple band structure effect.

Beatings in the amplitude of single-electron resonances are often
observed in Qdot structures and attributed to interference
modulation due to weak disorder. Indeed, as seen in the $dI/dV$
plot of Fig.~2b, the diamonds do not alternate regularly. In the
resonant tunneling model, one expects each single particle peak
to contribute negatively to the $TMR$ at sufficiently low
temperature. However, as we have measured the $TMR$ at
\mbox{$T=1.85$\,K}, where the single-particle resonances are
already strongly averaged out, the $TMR$ is only sensitive to the
average over these peaks, yielding a modulation that follows the
envelope function of the single-electron peaks.

The final proof that interference of single particle levels is the
physical origin for the observed $TMR$ oscillation comes from
measurements on single-wall carbon nano\-tubes (SWNTs). Figure~3
displays the conductance $G$ and the $TMR$ of a SWNT device. The
Qdot behaviour is already observed at \mbox{$1.85$\,K}, whereas
this was only evident at \mbox{$0.3$\,K} in the MWNT device. This
is consistent with the higher energy scales (both single-electron
charging energy and level spacing) for SWNTs as compared to MWNTs.
As seen in the $dI/dV$ plot of Fig.~3b, the typical
single-electron addition energy amounts to \mbox{$\sim 5$\,meV},
whereas it was an order of magnitude smaller in the MWNT device.

In Fig.~3a, the variation of the linear conductance
$G$ and the $TMR$ is simultaneously shown for two reso\-nances.
First, we observe that the $TMR$ changes sign on each conductance
re\-so\-nance. Furthermore, we see that the line shape of the
conductance resonances is sym\-metric, whereas that of the $TMR$ dips is asymmetric.
The jump in the $G(V_g)$ data at \mbox{$V_g=4.325$}
is not reproducible, but arises from background charge switching.
This jump is absent in a second scan.
% Note, that on the first resonance, $G$ displays a spurious switching.
% Such switchings, which are also observed in semiconducting Qdots, are likely caused
% by trapped charges in the oxide. They render the simultaneous
% measurement of $G$ and the $TMR$ difficult. On this example, the
% number of gate switches was low enough to allow a meaningful
% comparison between the two.
The amplitude of the $TMR$ ranges from
\mbox{$-7$\,\%} to \mbox{$+17$\,\%}, which is a higher amplitude
than for the MWNT samples. We think that this might be due to the higher
charging energy in SWNTs~$^{25}$.
Control experiments on two Normal-SWNT-Ferromagnetic \mbox{(N-SWNT-F)} devices
yield an order of magnitude lower signal
(see methods and supplementary online material),
proving that the current in the F-tube-F devices is indeed spin polarised.

The transmission probability through a Qdot near a resonance can
be described by the Breit-Wigner formula. If the Qdot is coupled
to two continua with spin-dependent densities of states, the
life-time of an electron on the Qdot becomes spin dependent.
Therefore, the width of the resonance is different for carriers
with up and down spins.
% The position in energy of an
% eigenstate in a carbon-nano\-tube Qdot, which is a one-dimensional
% Qdot, is determined by the round-trip phase acquired by an
% electron traveling in the nano\-tube and reflecting at its
% boundaries. Because the phases $\phi_{\sigma}$ of the reflection
% amplitudes depend on spin $\sigma$ if the contacts are
% ferromagnetic, the eigenstates will depend on the relative
% orientation of the magnetisation in the
% leads~\cite{27,28}.
% This is the conceptional basis of the mixing conductance introduced recently
% in spintronics for spin transport with non-collinear magnetisation.
In addition, the energy levels $E_n$ of the carbon-nano\-tube Qdot, acquire a spin-dependent part
caused by phases of the reflection amplitudes at the boundaries between the Qdot and the
ferromagnetic electrodes, which are spin dependent~$^{26-28}$.
% as seen from the resonance
% condition $2k_{n}L+\phi_{\sigma}^{L}+\phi_{\sigma}^{R}=2\pi n, n\in \mathbb{Z}$,
% where $\phi_{\sigma}^{L,R}$ denotes the phase change upon reflection of an electron
% with spin $\sigma$ on the left ($L$) or right ($R$) electrode.
The spin-dependent Breit-Wigner transmission probability for electrons
at energy $E$ with spin orientation $\sigma$ can conveniently be written as:
\begin{equation}
\label{eq:spinfet}
%\begin{displaymath}
T_{\sigma}=\frac{\Gamma_{L}^{\sigma}\Gamma_{R}^{\sigma}}{(E-E_{0}^{\sigma})^2+(\Gamma_{L}^{\sigma}+\Gamma_{R}^{\sigma})^2/4}
%\end{displaymath}
\end{equation}
where $\Gamma_{L(R)}^{\sigma}=\gamma_{L(R)}(1+\sigma P_{L(R)})$ denote
the spin-dependent \mbox{($\sigma = \pm 1$)} coupling to the left (right) ferromagnetic
lead, and $E_{0}^{\sigma}$ the spin-dependent energy level of the Qdot.
Note, that the polarisation in the leads $P_{L(R)}$ is measured
relative to the spin quantisation axis.

The sign change of the $TMR$ can be predicted with
Eq.~(\ref{eq:spinfet}), provided the couplings to the leads are
asymmetric, see Fig.~4. Off resonance, i.e. if $|E-E_{0}|\gg
(\Gamma_{L}^{\sigma}+\Gamma_{R}^{\sigma})$, $T_{\sigma}$ is small
and $\propto \Gamma_{L}^{\sigma}\Gamma_{R}^{\sigma}$, yielding the
normal positive $TMR$ of $+2P^2/(1-P^2)$ (we assume that
$|P_{L}|=|P_{R}|=P$). On resonance, on the other hand,
$T_{\sigma}\propto \Gamma_{L}^{\sigma}/\Gamma_{R}^{\sigma}$, (if
e.g. $\Gamma_{R}^{\sigma} \gg \Gamma_{L}^{\sigma}$), yielding an
anomalous negative $TMR$ of $-2P^2/(1+P^2)$. This mechanism has
already been suggested to explain an observed anomalous $TMR$ in
Ni/NiO/Co nanojunctions~$^{29}$. Unlike this earlier work, we are
able to follow the conductance and the $TMR$ by tuning the energy
level $E_0$ with the gate-voltage $V_g$ and compare with the
model. Whereas the negative $TMR$ can be understood following this
line of argument, the explicit shape and in particular the
asymmetry in the $TMR$ requires a spin-dependent energy level
$E_0^{\sigma}$ as we will show now. The eigenstate depends on the
gate-voltage \textit{and} on the spin direction:
$E_{0}^{\sigma}=E_{0}-\epsilon_{\sigma}-\alpha eV_{g}$, where
$\alpha$ is a constant proportional to the gate capacitance, $e$
the unit of charge, and $\epsilon_\sigma$ the spin-dependent part
of the energy level.
% It can be calculated from a Fabry-Perot like non-interacting one-dimensional scattering
% problem with tunnel barriers. Its value and functional form
% is sensitive to the interface considered, as well as to
% electron-electron interactions \cite{Egger:01,29}.
In the limit of small spin polarisation $P_{L(R)}\ll 1$, one may use the ansatz
$\epsilon_{\sigma}=\kappa\sigma(P_{L}+P_{R})$. We treat
$\kappa$ as a fitting parameter, which will be deduced from the experiment.
$\kappa$ determines the asymmetry of the $TMR$ signal.

The solid lines in Fig.~3 show fits to the measured
conductance $G$ and the $TMR$ using Eq.~(\ref{eq:spinfet}). As the
two resonances are well separated in energy, it is possible to fit
them individually. In order to obtain the conductance at finite
temperature, we convolved $T_{\sigma}$ with the derivative of the
Fermi-Dirac distribution at \mbox{$1.85$\,K} and sum over $\sigma$.
The following parameters entered the fits: $P=0.2$
% (this is slightly higher than the expected spin polarisation in our alloy of
% $0.1-0.15$, indicating that charging effect may amplify the spin signal)
and \mbox{$\kappa=0.32$\,meV}. $\gamma_{L,R}$ differ for the two
resonances: $\gamma_{L}=0.014$ \mbox{$(0.028)$\,meV} and
$\gamma_{R}=0.5$ \mbox{$(0.85)$\,meV} for the left (right)
resonance, respectively. Using these parameters, a very good
agreement between theory and experiment is found. Convincing
evidence for spin injection in a Qdot is deduced from the observed
asymmetric line shape of the $TMR$ in the SWNT device. The
spin-imbalance expressed by \mbox{$\epsilon_{\sigma}$} is
substantial, amounting to as much as \mbox{$\pm 0.13$\,meV} which
corresponds to an internal `exchange field' of \mbox{$B=2.2$\,T}.

% In summary, we have demonstrated a gate-tunable magnetoresistance effect
% in carbon nano\-tubes with PdNi-based ferromagnetic contacts. The
% $TMR$ oscillates as a function of a gate voltage between
% \mbox{$-5$\,\%} and \mbox{$+6$\,\%} for the MWNT and between
% \mbox{$-7$\,\%} to \mbox{$+17$\,\%} for the SWNT device. The observed
% spin-dependent phenomenon can fully be accounted for in a
% resonant tunneling picture.

% ===============================================================================================
% METHODS
% ===============================================================================================
\clearpage
\setlength{\parindent}{0mm}

{\small
  {\bf METHODS}\\
  {\bf Experimental}
  We have developed a reliable scheme to prepare relatively
  transparent ferromagnetic contacts to MWNTs and SWNTs using the
  ferromagnetic alloy Pd$_{1-x}$Ni$_{x}$ with $x\sim 0.7$~$^{20}$.
  Applying standard lithography techniques, a
  single nano\-tube is connected to two ferromagnetic PdNi
  electrodes that are further connected to bonding pads via Pd
  wires. We take advantage of the very good contacting properties of
  Pd to nano\-tubes~$^{30}$
  % \cite{Dai:03,Babic:04}
  and its giant paramagnetism.
  % \cite{Beille:75}
  We have studied and observed the $TMR$ on $9$ samples
  ($7$ MWNTs and $2$ SWNTs) with various tube lengths $L=0.4$, $0.5$, $0.8$, and \mbox{$1$\,$\mu$m}
  between the ferromagnetic electrodes. Control samples with Pd and PdNi contacts
  were fabricated as well, see below under `Control Experiment'.
}

{\small
  {\bf Stray-field effect}
  In order to rule out a simple stray-field effect, we compare the
  high-field magneto-resistance $S$ (defined as a $\%$ change of the
  resistance per T) with the low-field hysteretic $TMR$ signal.
  First, we see that the magnitude of the $TMR$ signal
  of the curves shown in Fig.~1. is to a good approximation constant
  (to \mbox{$3.4 \pm 0.4$\,\%}),
  whereas the background (high-field) magneto-resistance may change
  by as much as an order of magnitude.
  Secondly, the sign change of the $TMR$ from a
  positive value at \mbox{$V_g=-3.1$\,V} to a negative one at
  \mbox{$V_g=-3.3$\,V} is not accompanied by a change in the
  background $S$. In fact, all possible sign combinations of $S$ and
  $TMR$ have been observed. Because the low-field $TMR$ signal bears no
  correlation with the background magneto-resistance, we can exclude a
  stray-field effect from the contacts to the bulk nano\-tube as the
  source of the observed hysteretic signal.
  % Next, the detailed evolution of the $TMR$ with gate voltage $V_g$ is analysed.
}

{\small
  {\bf Control Experiment}
  In order to ensure that the measured $TMR$ is caused by a coherent
  spin polarised current, we have also analysed two devices with asymmetric contacts.
  One contact (F) is made from the ferromagnetic PdNi alloy of similar composition
  as it was used in the F-SWNT-F devices and the other (N) from the non-ferromagnetic metal Pd.
  This yields a N-SWNT-F device, which ideally should display no
  hysteretic signal.  Based on the noise signal of the resistance measurement
  a hysteretic switching signal (if any) must be smaller than \mbox{$1-1.5$\,\%}
  (supplementary online material, Fig.~S1).
  Because this is up to $10$ times smaller than
  what we have observed in the F-SWNT-F device for similar
  conductances, any magnetic artefact arising from a single ferromagnetic
  contact alone must be small, proving that we have observed a spin effect
  in transport in the F-SWNT-F case.
}

{\small\bf Received xx June 2005}

% ===============================================================================================
% REFERENCES
% ===============================================================================================
\clearpage

{\bf References}

% \begin{thebibliography}{33}
\begin{enumerate}

% REVIEW
% \bibitem{Wolf:2001}
\item
Wolf, S. A., Awschalom, D. D. \& Buhrman, R. A. \textit{et al.}
% Spintronics: a spin-based electronics vision for the future.
\textit{Science} {\bf 294}, 1488-95 (2001).

% REVIEW
% \bibitem{Zorpette:2001}
\item
Zorpette, G.
% The quest for the SPIN transistor.
\textit{IEEE Spectrum} {\bf 39}, 30-5 (2001).

% REVIEW
% \bibitem{Zutic:04}
\item
Zutic, I., Fabian, J. \& Das Sarma, S.
% Spintronics: Fundamentals and applications
\textit{Rev. Mod. Phys.} {\bf 76}, 323-410 (2004).

% TMR
% \bibitem{Juillere:75}
\item
Julli\`ere, M.
% Tunneling between ferromagnetic films.
\textit{Phys. Lett.} {\bf 54A}, 225-6 (1975).

% TMR
% \bibitem{Slonczewski:89}
\item
Slonczewski, J. C.
% Conductance and exchange coupling of two ferromagnets separated by a tunneling barrier.
\textit{Phys. Rev. B} {\bf 39}, 6995–7002 (1989)

% TMR
% \bibitem{Moodera:95}
\item
Moodera, J. S., Kinder, L. R., Wong, T. M. \& Meservey, R. Large
% Magnetoresistance at Room Temperature in Ferromagnetic Thin Film
Tunnel Junctions \textit{Phys. Rev. Lett.} {\bf 74}, 3273-6
(1995).

% GMR
% \bibitem{Baibich:88}
\item
Baibich, M. N., Broto, J. M., Fert, A., Nguyen van Dau, F. \& Petroff, F.
% Giant Magnetoresistance of (001)Fe/(001)Cr Magnetic Superlattices.
\textit{Phys. Rev. Lett.} {\bf 61}, 2472-5 (1988).

% GMR
% \bibitem{Binasch:89}
\item
Binasch, G., Gr{\"u}nberg, P., Saurenbach, F. \& Zinn, W.
% Enhanced magnetoresistance in layered magnetic structures with antiferromagnetic interlayer exchange.
\textit{Phys. Rev. B} {\bf 39}, 4828-30 (1989).

% INVERSE GMR
% \bibitem{George:94}
\item
George, J. M., Pereira, L. G., Barthelemy, A. \textit{et al.}
% Inverse spin-valve-type magnetoresistance in spin engineered multilayered structures.
\textit{Phys. Rev. Lett.} {\bf 72}, 408-11 (1994).

% SPIN FET
% \bibitem{Datta:90}
\item
Datta, S. \& Das, B.
% Electronic analog of the electro-optic modulator.
\textit{Appl. Phys. Lett.} {\bf 56}, 665-7 (1990).

% SPIN FET
% \bibitem{Schaepers:00}
\item
Sch{\"a}pers, Th., Nitta, J., Heersche, H. B. \& Takayanagi, H.
% Interference ferro\-magnet-semi\-con\-duc\-tor-ferro\-magnet spin field-effect transistor.
\textit{Phys. Rev. B} {\bf 64}, 125314 (2000).

% TMR in MWNTs
% \bibitem{Tsukagoshi:99}
\item
Tsukagoshi, K., Alphenaar, B. W. \& Ago, H.
% Coherent transport of electron spin in a ferro\-magnetically contacted carbon nanotube.
\textit{Nature} {\bf 401}, 572-4 (1999).

% Spin polarization in tunnelling
% \bibitem{Tedrow:73}
\item
Tedrow, P. M. \& Meservey, R.
% Spin Polarization of Electrons Tunneling from Films of Fe, Co, Ni, and Gd.
\textit{Phys. Rev. B} {\bf 7}, 318–26 (1973).

% TMR in MWNTs
% \bibitem{Schneider:00}
\item
Zhao, B., M\"{o}nch, I., Vinzelberg, H., M\"{u}hl, T. \& Schneider, C. M.
% Spin-coherent transport in ferro\-magnetically contacted carbon nanotubes.
\textit{Appl. Phys. Lett.} {\bf 80}, 3144-6 (2002).

% \bibitem{Tans:97}
\item
Tans, S. J., Devoret, M. H., Dai, H., Thess, A., Smalley, R. E., Geerligs, L. J. \& Dekker, C.
% Individual single-wall carbon nano\-tubes as quantum wires.
\textit{Nature} {\bf 386}, 474-7 (1997).

% \bibitem{Bockrath:97}
\item
Bockrath, M., Cobden, D. H., McEuen, P. L., Chopra, N. G., Zettl, A., Thess, A. \& Smalley, R. E.
% Single-Electron Transport in Ropes of Carbon Nanotubes.
\textit{Science} {\bf 275}, 1922-5 (1997).

% \bibitem{Liang:01}
\item
Liang, W., Bockrath, M., Bozovic, D., Hafner, J. H., Tinkham, M. \& Park, H.
% Fabry - Perot interference in a nanotube electron waveguide.
\textit{Nature} {\bf 411}, 665-9 (2001).

% \bibitem{Dai:01}
\item
Kong, J., Yenilmez, E., Tombler, T. W., Kim, W. \& Dai, H.
% Quantum Interference and Ballistic Transmission in Nanotube Electron Waveguides.
\textit{Phys. Rev. Lett.} {\bf 87}, 106801 (2001).

% \bibitem{Buitelaar:02}
\item
Buitelaar, M. R., Bachtold, A., Nussbaumer, T., Iqbal, M. \& Sch{\"o}nenberger, C.
% Multiwall Carbon Nanotubes as Quantum Dots.
\textit{Phys. Rev. Lett.} {\bf 88}, 156801 (2002).

% \bibitem{Sahoo:05}
\item
Sahoo, S., Kontos, T., Sch{\"o}nenberger C. \& S{\"u}rgers, C.
% Electrical spin injection in multiwall carbon nanotubes with transparent ferromagnetic contacts.
\textit{Appl. Phys. Lett.} {\bf 86}, 112109 (2005).

% \bibitem{Rashba:84}
\item
Bychkov, Yu. A., Rashba, E. I.
% Oscillatory effects and the magnetic susceptibility of carriers in inversion layers.
\textit{J. Phys. C} {\bf 17}, 6039-45 (1984).

% \bibitem{Nitta:97}
\item
Nitta, J., Akazaki, T., Takayanagi, H. \& Enoki, T.
% Gate Control of Spin-Orbit Interaction in an Inverted In$_{0.53}$Ga$_{0.47}$As/In$_{0.52}$Al$_{0.48}$As Heterostructure.
\textit{Phys. Rev. Lett.}, {\bf 78}, 1335-8 (1997).

% \bibitem{Cobden:98}
\item
Cobden, D. H., Bockrath, M., McEuen, P. L., Rinzler, A. G. \& Smalley, R. E.
% Spin Splitting and Even-Odd Effects in Carbon Nanotubes.
\textit{Phys. Rev. Lett.} {\bf 81}, 681-4 (1998).

% \bibitem{Egger:05}
\item
De Martino, A. \& Egger, R.
% Rashba spin–orbit coupling and spin precession in carbon nanotubes.
\textit{J. Phys. Cond. Matter} {\bf 7}, 5523 (2005).

% \bibitem{Fert:00}
\item
Barnas, J., Martinek, J., Michalek, G., Bulka, B. R. \& Fert, A.
% Spin effects in ferromagnetic single-electron transistors.
\textit{Phys. Rev. B} {\bf 62}, 12363–73 (2000)

% \bibitem{Brataas:00}
\item
Brataas, A., Nazarov, Yu. V., \& Bauer, G. E. W.
% Finite-Element Theory of Transport in Ferromagnet–Normal Metal Systems.
\textit{Phys. Rev. Lett.} {\bf 84}, 2481-4 (2000).

% \bibitem{Waintal:00}
\item
Waintal, X., Myers, E. B., Brouwer, P. W. \& Ralph, D. C.
% Role of spin-dependent interface scattering in generating current-induced torques in magnetic multilayers.
\textit{Phys. Rev.} B {\bf 62}, 12317 (2000).

% \bibitem{Koenig:04}
\item
Braun, M., K\"onig, J. \& Martinek, J.
% Theory of transport through quantum-dot spin valves in the weak-coupling regime.
\textit{Phys. Rev. B} {\bf 70}, 195345 (2004).

% \bibitem{Tsymbal:98}
\item
Tsymbal, E. Y., Sokolov, A., Sabirianov, I. F. \& Doudin, B.
% Resonant Inversion of Tunneling Magnetoresistance.
\textit{Phys. Rev. Lett.} {\bf 90}, 186602 (2003).

% \bibitem{Dai:03}
\item
Javey, A., Guo, J., Wang, Q., Lundstrom, M. \& Dai, H.
% Ballistic carbon nanotube field-effect transistors
\textit{Nature} {\bf 424}, 654-7 (2003).

% \end{thebibliography}
\end{enumerate}

% ===============================================================================================
% END of REFERENCES
% ===============================================================================================

% ===============================================================================================
% Supplementary Information and Acknowledgement
% ===============================================================================================
{\noindent\small{\bf Supplementary Online Material} Figure~1S is available at ...}
\vspace{5mm}

{\noindent\small{\bf Acknowledgments}
  We acknowledge fruitful discussions with R. Allenspach, W. Belzig, R. Egger, and
  H. S. J. van der Zant. We thank L. Forr\'o for providing the MWNTs.
  This work has been supported by the EU RTN network DIENOW,
  the Swiss National Center (NCCR) on nano-scale science,
  and the Swiss National Science Foundation.
}
\vspace{5mm}

{\noindent\small{\bf Author Information} Reprints and permissions information
is available at\\
npg.nature.com/reprintsandpermissions. The authors declare no competing financial
interests. Correspondence and request for materials should be addressed to C.S. (christian.\-schoenenberger\-@unibas.ch)}

% ===============================================================================================
% Begin of FIGURES, actually only CAPTIONS
% ===============================================================================================

\newpage
%\begin{figure}[htb]
%\includegraphics[width=0.8\linewidth]{Fig1.eps}
%\end{figure}
\noindent {\bf Figure~1 $\mid$ The tunneling magnetoresistance TMR changes sign with gate voltage $V_g$.}\\
  Inset: SEM picture of a carbon nanotube contacted to ferromagnetic PdNi strips.
  The separation between the
  contacts along the nanotube amounts to \mbox{$L=400$\,nm}. The magnetic
  field $H$ was applied in plane. No qualitative difference has been seen
  for the field direction parallel and perpendicular
  to the long axis of the ferromagnetic electrodes.\\
  Main panel: Linear response resistance $R$
  as a function of magnetic field $H$ at \mbox{$T=1.85$\,K} for different
  gate voltages $V_g$. The blue (red) arrow indicates the up (down)
  magnetic field sweep direction, respectively. The observed
  amplitude and the sign of the $TMR$ depend on $V_g$, but not on the
  high field magnetoresistance (MR), which is expressed by $S$ denoting
  the percentage change of the MR with magnetic field (see methods).
  We note, that we extract the maximum possible value for the $TMR$ signal as
  indicated in the figure.

\clearpage
%\begin{figure}[htb]
%\includegraphics[width=0.08\linewidth]{Fig2.eps}
%\end{figure}
\noindent {\bf Figure~2  $\mid$ The TMR of the device shown in figure~1 oscillates with gate volte $V_g$.}\\
  {\bf a} $TMR$ as a function of gate voltage $V_g$
  at \mbox{$T=1.85$\,K}. The characteristic gate voltage scale
  \mbox{$\Delta V_g^{TMR}$} of the observed $TMR$ modulation
  varies between $0.4$ and \mbox{$0.75$\,V}.
  The bars reflect the error in deducing the TMR signal from $R(B)$ curves, see Fig.~1.\\
  {\bf b,c} This data was measured in a different cryostat at \mbox{$300$\,mK}:
  {\bf b} non-linear differential conductance $dI/dV$ as
  a function of source-drain $V_{sd}$ and gate voltage $V_g$ in a narrow
  $V_g$ interval, corresponding to the addition of $6$ electrons; and
  {\bf c} the linear conductance $G$ over a much wider $V_g$ interval.
% The conductance beatings and the corresponding TMR oscillations
% can be captured in a simple multichannel transport model.
% Mode coupling needs to be consider to account for the observation.
% See, also Ref.~\cite{Krompiewski:04}.

\clearpage
%\begin{figure}[htb]
%\includegraphics[width=0.08\linewidth]{Fig3.eps}
%\end{figure}
\noindent {\bf Figure~3 $\mid$ The detailed evolution of the TMR signal for two single-particle resonances.}\\
  Results for a SWNT device with a contact
  separation of \mbox{$L=500$\,nm} measured at \mbox{$T=1.85$\,K}.\\
  {\bf a} Measurement ($\blacksquare$) of the linear conductance $G$ and the $TMR$
  around two resonances, and theoretical fit (solid curves) using Eq.~(\ref{eq:spinfet})
  with $P=0.2$, \mbox{$\kappa = 0.32$\,meV} and \mbox{$\gamma_{L}=0.014\,(0.028)$\,meV}
  and \mbox{$\gamma_{R}=0.5\,(0.85)$\,meV} for the left\,(right) resonance, respectively.
  The bars in Fig.~3a reflect the error in deducing the TMR signal from $R(B)$ curves.\\
  {\bf b} Plot of the non-linear differential conductance $dI/dV$ as
  a function of source-drain $V_{sd}$ and gate voltage $V_g$.
  Typical $R(B)$ curves for this F-SWNT-F are shown in Fig.~S1 (supplementary online material),
  as are two representative $R(B)$ curves for N-SWNT-F devices.

\clearpage
%\begin{figure}[htb]
%\includegraphics[width=0.08\linewidth]{Fig4.eps}
%\end{figure}
\noindent {\bf Figure~4 $\mid$ Schematics explaining the observed sign change of the TMR. }\\
  {\bf a} Depicts the case for antiparallel and
  {\bf b} for parallel magnetisation.
  The electric resistance of an asymmetric resonant-tunneling junction
  (we assume $\Gamma_R >> \Gamma_L$) is proportional to the asymmetry $A:=\Gamma_R/\Gamma_L$,
  where $\Gamma_{R,L}$ are the tunneling rates to the right
  (R) and left (L) electrode, respectively.
  If ferromagnetic contacts are used, $\Gamma_{R,L}$ become spin-dependent.
  The rate is increased for the spin direction
  of the majority carriers, whereas it is decreased for the minority ones. The two spin-directions
  are colour-coded (red $=$ up spin and blue $=$ down spin). Because the electric resistance
  is spin-dependent, the total resistance $R$ is the parallel circuit
  of $R^{\uparrow}$ and $R^{\downarrow}$. Whereas $R^{\uparrow}$ and $R^{\downarrow}$
  are equal in the parallel configuration, $R^{\uparrow}$ is smaller and $R^{\downarrow}$ is larger
  in the antiparallel configuration.
  Due to the dominance of the smaller resistance in a parallel circuit,
  $R$ is smaller in the antiparallel as compared to the parallel case, corresponding to a
  \emph{negative} $TMR$ signal.

% ===============================================================================================
% END of DOCUMENT
% ===============================================================================================

\end{document}